\documentclass[useAMS,referee,usenatbib]{biom}
\usepackage[pdftex]{graphicx}
\usepackage{subcaption}
\usepackage[font=small,labelfont=bf]{caption} 
\usepackage{amsmath,amssymb}  
\usepackage{algorithm}
\usepackage{algpseudocode}
\usepackage{epsfig}  
\usepackage{setspace}
\usepackage{xcolor}
\usepackage{siunitx}
\usepackage{url}
\usepackage{hyperref}
\def\bSig\mathbf{\Sigma}

\title[Graph-Adaptive Horseshoe for Compositional Regression]{Graph-Adaptive Horseshoe for Compositional Regression}

\author{Satabdi Saha\email{ssaha1@mdanderson.org} \\
Department of Biostatistics, The University of Texas MD Anderson Cancer Center, Houston, TX, USA
\and
Christine B. Peterson\emailx{cbp2@rice.edu} \\
Department of Statistics, Rice University, Houston, TX, USA}

\usepackage[table]{xcolor}
\begin{document}


\date{{\it Received } 2025. {\it Revised } 2025.  {\it
Accepted } 2025.}



\pagerange{\pageref{firstpage}--\pageref{lastpage}} 
\volume{1}
\pubyear{2025}
\artmonth{September}


\doi{10.1111/j.1541-0420.2005.00454.x}


\label{firstpage}


\begin{abstract}
 Compositional predictors, such as microbiome abundances, pose unique challenges in variable selection due to their unit-sum constraint and inherent dependencies. 
Existing approaches often rely on fixed association graphs derived from phylogenetic 
or ecological distances, which may not reflect outcome-relevant relationships. 
We propose GRACE (GRaph-Adaptive horseshoe for Compositional rEgression), 
a fully Bayesian framework that enforces compositional constraints, performs variable 
selection, and adaptively learns an outcome-driven shrinkage graph. GRACE achieves 
compositionality through a novel linear reparameterization of regression coefficients, 
while a structured horseshoe prior induces sparsity and smooths coefficients along the 
learned graph. Graph learning is accomplished via scaled beta2 priors on edge weights, 
providing both outcome-specific adaptation and posterior uncertainty quantification. 
We develop an efficient Gibbs sampler incorporating elliptical slice sampling to ensure scalability in high dimensions. Through extensive simulations, GRACE demonstrates competitive predictive accuracy 
and improved graph recovery compared with existing methods, particularly under 
graph misspecification. \textcolor{blue}{Application to oral microbiome data from the ORIGINS study identifies taxa associated with insulin resistance and yields an outcome-driven graph summarizing how those taxa relate to the outcome, a structure that differs substantially from phylogenetic or co-occurrence networks.} These findings highlight that fixed predictor graphs useful for regularization may not faithfully represent outcome-relevant feature relationships, underscoring the need for adaptive, outcome-informed approaches in compositional regression.


\end{abstract}

%



\maketitle


%

\section{Introduction}
\label{sec:Introduction}

Compositional data arise naturally in many fields, including geology, ecology, economics, and notably, microbiome studies. A distinctive characteristic of these datasets is that the observed values represent relative abundances, meaning that the feature values for each sample represent proportions of a whole rather than absolute measures. As a result, the components within each sample must sum to one, placing the data on a simplex and introducing inherent dependencies among features. Consequently, traditional regression approaches become inappropriate when compositional data serve as predictors \citep{lin2014variable,peterson2023analysis}. Human microbiome data pose additional challenges: they are often high-dimensional and exhibit phylogenetic structure, meaning that microbial taxa are organized according to evolutionary relationships captured in a phylogenetic tree.
Incorporating such structure into predictive models can improve accuracy and interpretability, particularly when evolutionary relatedness aligns with functional similarity or outcome relevance.


To handle compositional predictors in regression, 
\cite{aitchison1984log} proposed
the log-contrast regression model, which transforms the constrained compositional predictors into an unconstrained Euclidean space, thereby enabling accurate estimation of regression coefficients.
 To extend this framework to microbiome datasets, \cite{lin2014variable} introduced an $l_1$-penalized version of the linear log-contrast model, allowing for sparse coefficient estimation and improved predictive performance in high-dimensional settings. However, these methods do not exploit the inherent  structure among microbial taxa. 
 To incorporate such structure, Liangliang Zhang et al.\ (\citeyear{zhang2021bayesian}) proposed a Bayesian compositional regression model that encodes the compositionality constraint through the prior covariance of the coefficient vector and uses an Ising prior to encourage joint selection of phylogenetically related features. However, this approach does not provide continuous shrinkage for nonzero coefficients and remains sensitive to the fixed similarity matrix used in the Ising prior.
 More recently, Li Zhang et al.\ (\citeyear{zhang2024bayesian}) developed a Bayesian compositional generalized linear model, which incorporates phylogenetic relatedness through a structured regularized horseshoe prior. 
 Although this model introduces adaptive shrinkage, it still inherits its neighborhood structure completely from a user‑supplied similarity matrix. 
\textcolor{blue}{For both models, the fixed similarity structure does not allow the graph to adapt to the outcome, making it difficult to distinguish similarity induced by the supplied graph from similarity supported by the outcome data.} In \citep{saha2024bayesian}, we introduced a Bayesian compositional regression model achieving feature aggregation by clustering the regression coefficients, resulting in a flat grouping of predictors. 
\textcolor{blue}{In contrast, this work moves from flat coefficient clustering to graph learning. We aim to estimate an outcome-informed similarity graph among predictors that represents related taxa as connected effect-sharing modules.
.
}


To motivate our proposed approach, we analyze data collected under the  Oral Infections, Glucose Intolerance, and Insulin Resistance Study  \citep[ORIGINS,][]{demmer2015periodontal, demmer2017subgingival}, which seeks to characterize the association between the bacterial population of subgingival plaque and prediabetes in adults. Oral microorganisms play a key role in shaping the risk of gum diseases, including periodontitis, a serious infection of the soft tissue around the teeth \citep{pihlstrom2005periodontal}.
It is postulated that chronic inflammation driven by the periodontal microbiota may contribute to impaired glucose regulation and heightened risk of insulin resistance, potentially laying the groundwork for type 2 diabetes \citep{gurav2012periodontitis}. The ORIGINS study enrolled 152 diabetes-free individuals (77\% female), aged between 20 and 55 years. Subgingival plaque samples were profiled using the Human Oral Microbe Identification Microarray (HOMIM) \citep{colombo2009comparisons}, yielding abundance data for 379 microbial taxa. Our aim is to identify taxa associated with insulin resistance, thereby elucidating the influence of the periodontal microbiome on prediabetes.  We use  HOMA-IR (Homeostatic Model Assessment of Insulin Resistance), defined as
\[
\mathrm{HOMA\text{-}IR}
\;=\;
\frac{\bigl[\text{fasting insulin (\si{\micro U/mL})}\bigr]\times\bigl[\text{fasting glucose (\si{mmol/L})}\bigr]}{22.5}\,,
\]
as a validated index of whole-body insulin resistance that can be obtained from a single fasting blood draw \citep{ demmer2017subgingival}.  Because insulin resistance typically precedes overt hyperglycemia by years, HOMA-IR serves as an early, continuous marker of prediabetic metabolic dysfunction. We use \(\log(\text{HOMA-IR}\)) as the outcome because HOMA-IR is typically right-skewed, and the log transformation reduces upper-tail influence and better supports linear-model assumptions.

\begin{figure}
    \centering
    \includegraphics[width=1.0\linewidth]{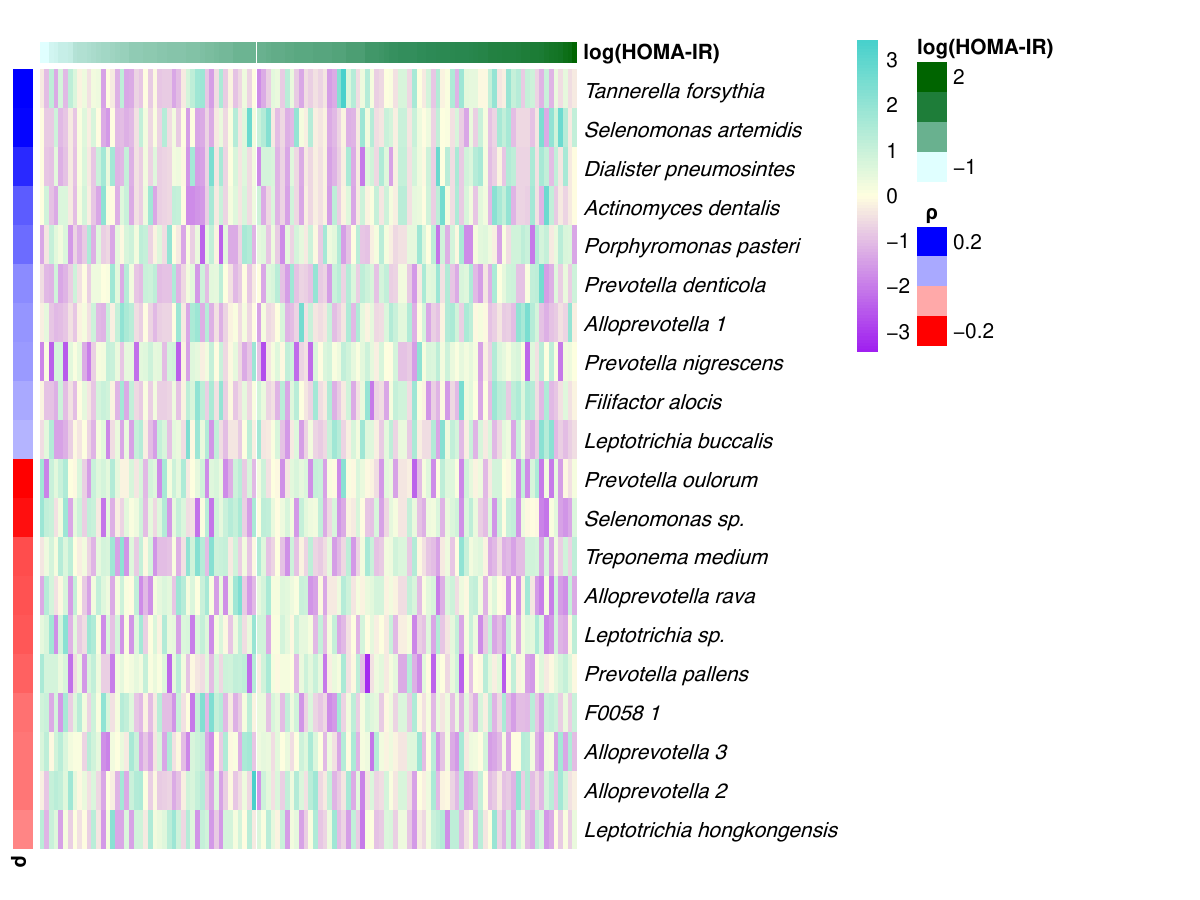}
    \caption{Heatmap of the 20 taxa showing the strongest univariate association with insulin resistance. Rows correspond to species; columns are the 118 periodontal samples, ordered left-to-right by increasing log(HOMA-IR). The CLR-transformed abundance values are row-standardized: yellow indicates the row mean, blue shades represent abundances above the mean, and red shades represent abundances below the mean. The vertical color strip on the left encodes the Spearman correlation coefficient ($\rho$) between each taxon’s abundance and log(HOMA-IR): blue denotes positive $\rho$, and red denotes negative $\rho$. The horizontal bar above the heatmap displays each sample’s log(HOMA-IR). Together, the plot highlights taxa that tend to be enriched (blue $\rho$ values) or depleted (red $\rho$ values) as insulin resistance increases.}
    \label{fig:heatmap}
\end{figure}
\begin{figure}
    \centering
    \includegraphics[width=1\linewidth]{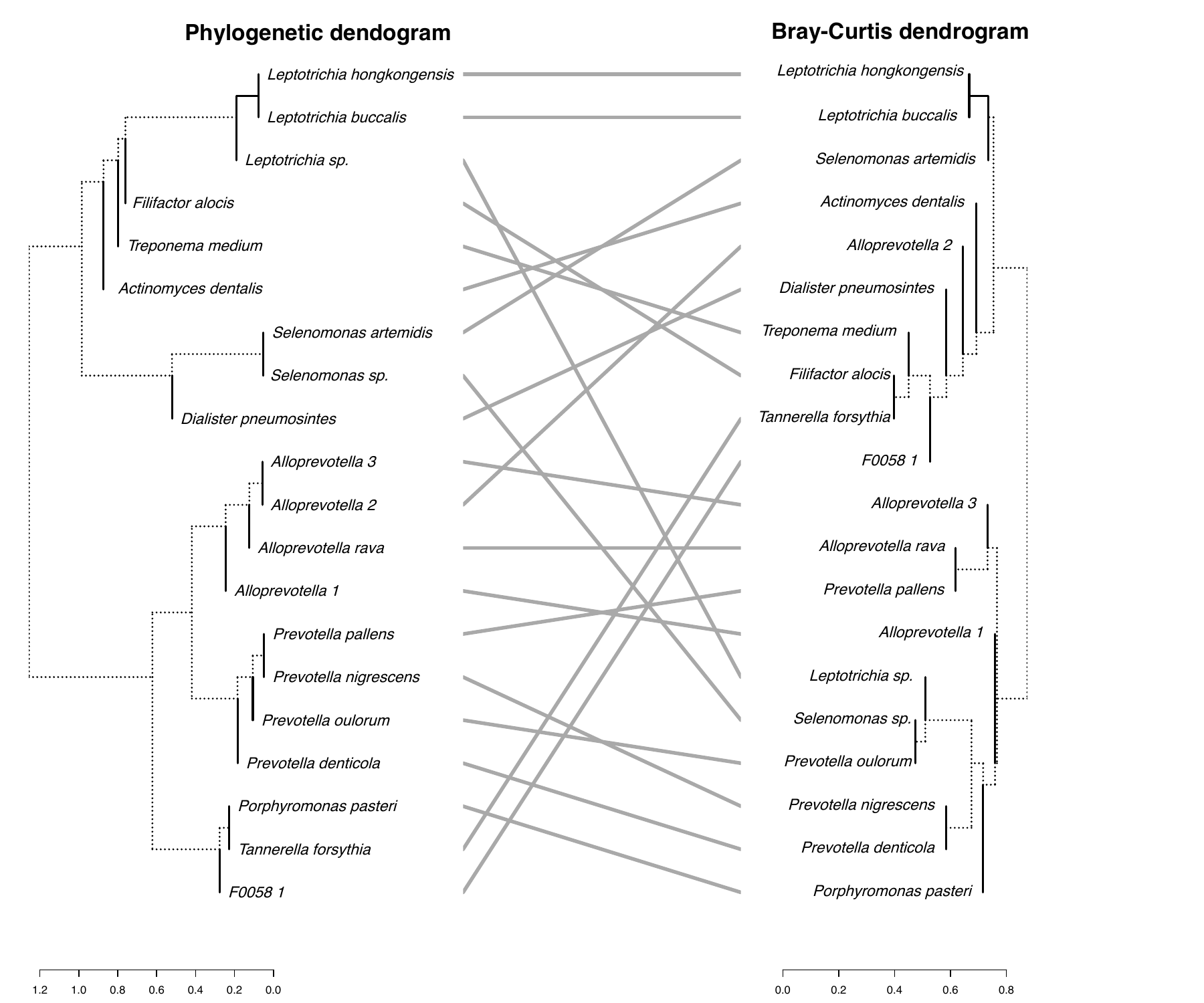}
    \caption{Tanglegram of the 20 taxa most strongly associated with insulin resistance (10 highest positive and 10 highest negative Spearman $\rho$ vs. log HOMA‑IR) as reported in Figure \ref{fig:heatmap}. The left dendrogram shows a tree obtained from hierarchical clustering by normalized phylogenetic (cophenetic) distances, while the right dendrogram shows clustering by Bray–Curtis dissimilarity on the CLR transformed abundances. Gray lines connect each taxon’s position in the two trees; extensive crossings indicate that taxa sharing common ancestry often occupy distinct ecological modules, and conversely that co‑occurrence patterns in relation to HOMA-IR cut across deep evolutionary clades.}
    \label{fig:Tanglegram}
\end{figure}

To assess whether individual taxa carry predictive information for insulin resistance, we computed Spearman’s rank correlations \(\rho\) between centered log ratio (CLR) transformed taxon abundances and \(\log(\text{HOMA-IR})\), and plotted the top 20 positive and negative associations (Figure~\ref{fig:heatmap}). Several taxa showed moderate positive or negative correlations, motivating the subsequent multivariable compositional regression analysis. To characterize inter-taxon relationships, one might impose a fixed weight matrix, such as a phylogenetic covariance matrix or Bray--Curtis co-abundance graph. However, these microbial relatedness measures often disagree (Supplementary Figures S8 and S9). Figure~\ref{fig:Tanglegram} compares the phylogenetic tree and Bray--Curtis dendrogram for the same 20 taxa, showing that taxa close in evolutionary distance, such as \emph{Prevotella} spp., are not necessarily close in compositional space, and vice versa. Moreover, neither structure incorporates relevance to insulin resistance: the warm and cool blocks in Figure~\ref{fig:heatmap} diverge from both dendrograms. \textcolor{blue}{These discrepancies motivate learning an outcome-informed shrinkage graph that links taxa through shared regression effects, rather than relying on a single fixed taxa-only distance.}

We propose GRACE (GRaph-Adaptive horseshoe for Compositional rEgression), 
a fully Bayesian hierarchical model for compositional regression that jointly performs 
variable selection and learns an outcome-informed shrinkage graph from the data. 
Unlike methods that rely on fixed external structures, GRACE allows the graph to adapt 
to effect-sharing patterns supported by the data, yielding interpretable groups of taxa 
with similar outcome associations. Our framework introduces two key innovations. First, it enforces the compositional 
constraint \(\sum_{j=1}^p \beta_j=0\) through a linear reparameterization of 
\(\boldsymbol{\beta}\), enabling efficient posterior inference. Second, it uses a shrinkage prior to promote sparse variable selection while directly learning the shrinkage graph, thereby providing posterior uncertainty quantification for the graph.
To make inference feasible in high dimensions, we further develop a customized Gibbs sampler 
incorporating elliptical slice sampling to improve mixing and scalability.

The remainder of the article is organized as follows. Section~2 introduces GRACE and the MCMC algorithm. Sections~3 and~4 present the simulation design and benchmarking results. Section~5 applies GRACE to the ORIGINS oral microbiome data, and Section~6 concludes with a discussion and future directions.

\section{Methods}
\label{s:model}
\subsection{Background}
Our model builds on the compositional regression framework for predicting continuous outcomes from microbiome profiles. Let \(\boldsymbol{y}=\{y_1,\ldots,y_n\}\) denote the vector of continuous responses for \(n\) subjects, and let \(\mathbf{M}\) be the \(n\times p\) observed count matrix for \(p\) microbial features. Because microbial counts are compositional and interpretable only on a relative scale, we convert counts to relative abundances using total sum scaling (TSS), typically after adding a small pseudocount to avoid numerical instability from zeros \citep{lin2014variable,bien2021tree}. The resulting matrix \(\tilde{\mathbf{M}}\) satisfies \(\sum_{j=1}^p \tilde{m}_{ij}=1\) for each subject \(i\), so each row lies on the simplex \(\mathcal{S}^p\) rather than unrestricted Euclidean space \(\mathbb{R}^p\). We then define the log-relative abundance matrix \(\mathbf{X}=\log(\tilde{\mathbf{M}})\). However, despite this transformation, the features in $\mathbf{X}$ remain statistically dependent due to the underlying compositional constraint.

\subsection{Model specification}

\textcolor{blue}{We consider the linear log-contrast model for compositional regression \citep{lin2014variable}}:
\begin{align}
\label{eq:constrainedLR}
    \boldsymbol{y} = \mathbf{X}\boldsymbol{\beta}+ \boldsymbol{\varepsilon} \hspace{0.1in} \quad \text{subject to the constraint} \hspace{0.1in} \boldsymbol{1}^{\top}\boldsymbol{\beta} = 0.
\end{align}
 To remove the intercept, both the response and compositional predictors are centered prior to modeling. \textcolor{blue}{This formulation is equivalent to the usual log-contrast model \cite{aitchison1984log} at the level of the induced linear predictor, while avoiding dependence on a specific reference component. Complete derivations are provided in Supplementary Section~1.}
 In the current work, we propose a novel Bayesian approach for addressing the high dimensionality and compositionality of $\mathbf{X}$ through a horseshoe-like shrinkage structure on the regression coefficient vector:
\begin{align} 
    \bm{\beta} &\sim \mathcal{N}(\bm{0}, \sigma^2 \zeta^2 \mathbf{U} ^{-1} \mathbf{\Lambda} (\mathbf{U} ^{-1} )^{\top}), \qquad   \mathbf{\Lambda} = \text{Diag}(\lambda_j ^2), \quad \mathbf{U}^{\top}\mathbf{U} =\mathbf{T}^{\top}\mathbf{T},  \nonumber\\
    \lambda_j \,&\sim\, \mathcal{C}^+(0,1), \qquad
  \zeta \,\sim\, \mathcal{C}^+(0,\zeta_0).
  \label{eq:beta-prior}
\end{align}
{\color{blue}Let \(\mathbf{1}_{p}\in\mathbb{R}^{p}\) represent the all-ones vector and $c$ a scalar $>0$. Then we can define $\mathbf{T} = \begin{bmatrix}
 \mathbf{I}_p \\
 c \times \bm{1}'_p
 \end{bmatrix} \in \mathbb{R}^{(p+1)\times p}$, 
\(\mathbf{A} = \mathbf{T}^{\top}\mathbf{T} = \mathbf{I}_{p} + c^{2}\,\boldsymbol{1}_{p}\,\boldsymbol{1}_{p}^{\top},
\)
and 
\(\mathbf{U} \;=\; \operatorname{chol}\bigl(\mathbf A\bigr)\).} Note that equation \eqref{eq:beta-prior} encodes the horseshoe shrinkage structure with $\zeta$ being the global scale parameter and $\lambda_j$ being the local shrinkage parameter for each predictor, $j = 1, \ldots, p$. Following the regularized horseshoe structure \citep{piironen2017sparsity},  we define $\zeta_0 = \frac{p_0}{p - p_0}\frac{\sigma}{\sqrt{n}}$, where $p_0$ is the prior guess for the number of non-zero predictors. Theorem 1 details how the prior formulation of $\boldsymbol{\beta}$ satisfies the compositionality constraint while retaining the horseshoe structure. The detailed proof is deferred to Section 2 of the Supplementary Materials. In addition, in Theorem 2, given in Section 5 of the Supplement, we prove that as $c \to \infty$, the posterior distribution of the linear functional ${\bf 1}^\top \boldsymbol \beta$ collapses to a point mass at zero, thereby enforcing the compositionality constraint.


\begin{theorem}
Let \(p\ge1\), and denote by \(\mathbf{1}_{p}\in\mathbb{R}^{p}\) the all-ones vector.  For each \(c>0\), define
\(
\mathbf{T}(c) \;=\;
\begin{bmatrix}
\mathbf{I}_{p} \\
c\,\boldsymbol{1}_{p}^{\!\top}
\end{bmatrix}
\;\in\;\mathbb{R}^{(p+1)\times p}, 
\quad
\mathbf{A}(c) \;=\; \mathbf{T}(c)^{\!\top}\mathbf{T}(c) \;=\; \mathbf{I}_{p} \;+\; c^{2}\,\boldsymbol{1}_{p}\,\boldsymbol{1}_{p}^{\!\top},
\)
and let
$
\mathbf{U}(c) \;=\; \operatorname{chol}\bigl(A(c)\bigr)
\quad\text{so that}\quad
\mathbf{U}(c)^{\!\top}\mathbf{U}(c) \;=\; \mathbf{A}(c).
$
Suppose \(\boldsymbol{\eta}\sim N_{p}\bigl(\boldsymbol{0},\,\sigma^{2}\zeta^2\mathbf{\Lambda}\bigr)\) for a fixed diagonal matrix 
\(\mathbf{\Lambda}=\operatorname{diag}(\lambda_{1}^{2},\dots,\lambda_{p}^{2})\) with \(\lambda_{j}>0\), and set
$
\boldsymbol{\beta}(c) \;=\; \mathbf{U}(c)^{-1}\,\boldsymbol{\eta} \;\in\;\mathbb{R}^{p}.
$
Then
\(
\operatorname{Var}\!\bigl(\mathbf{1}_{p}^{\!\top}\boldsymbol{\beta}(c)\bigr)
\;=\;
\sigma^{2} \zeta^2\,\boldsymbol{1}_{p}^{\!\top}\,\mathbf{U}(c)^{-1}\,\mathbf{\Lambda}\,\mathbf{U}(c)^{-\!\top}\,\boldsymbol{1}_{p}
\;\longrightarrow\;0
\quad\text{as }c\to\infty.
\)
In particular, as \(c\to\infty\), the prior variance of the sum \(\sum_{j=1}^{p}\beta_{j}(c)\) vanishes, so \(\sum_{j}\beta_{j}(c)\to0\) in probability, enforcing the compositionality constraint. 
\end{theorem}

\subsection{Learning the feature graph} 

We propose to the learn outcome-dependent feature associations by defining a graph on the local shrinkage parameters \(\lambda_j\), encouraging graph-connected taxa to have similar shrinkage and hence smoother regression effects. Since we learn the graph from the data, its posterior summaries capture outcome-informed effect-sharing relationships distinct from phylogenetic or co-occurrence networks.
To achieve this, we first define
\(\psi_j = \log(\lambda_j)\) as in Li Zhang et al.\ (\citeyear{zhang2024bayesian}) and 
place an intrinsic autoregressive prior \citep{besag1995conditional} 
on \(\boldsymbol{\psi}\), where $j \sim k$ represents a neighbor relation:   \[
p(\boldsymbol{\psi}) \propto \exp \left\{ -\frac{1}{2} \sum_{j \sim k} \frac{1}{\tau^2} \omega_{jk} (\psi_j - \psi_k)^2 \right\}.
\]
The weights \(\omega_{jk}\) represent the strength of the relationship between nodes \(j\) and \(k\), where \( \omega_{jk} = 0 \) if there is no connection and \( \omega_{jk} > 0 \) if there is a connection. 
The global variance parameter \(\tau^2\) controls the overall smoothness, and the sum of weights \(\omega_{jk}\) reflects the influence of the neighboring nodes on \(\psi_j\). Instead of fixing the weights \(\omega_{jk}\) to a predetermined similarity matrix, we propose to place a scaled beta\(2\)  prior \citep{perez2017scaled} elementwise on the weights:
{\color{blue}
\begin{align}
\label{eq:sb2}
\omega_{jk}\;&\sim\;\mathrm{SBeta2}\bigl(\alpha_{jk},\,q_{jk},\,\kappa_{jk}\bigr) \quad \alpha_{jk},\,q_{jk},\,\kappa_{jk} >0\\
f(\omega_{jk}\mid \alpha_{jk},q_{jk},\kappa_{jk}) 
&=
\frac{\Gamma(\alpha_{jk}+q_{jk})}
     {\Gamma(\alpha_{jk})\,\Gamma(q_{jk})\,\kappa_{jk}}
\Biggl(\frac{\omega_{jk}}{\kappa_{jk}}\Biggr)^{\alpha_{jk}-1}
\biggl(1 + \frac{\omega_{jk}}{\kappa_{jk}}\biggr)^{-(\alpha_{jk}+q_{jk})},
\quad
\omega_{jk}>0, \nonumber 
\end{align}}
where $\alpha_{jk},\,q_{jk},\,$ and $\kappa_{jk}$ are fixed hyperparameters. 
We adopt the scaled beta2 prior as this family of distributions provides a flexible framework for modeling the weights, with hyperparameters that can be tuned to capture both highly informative and nearly non-informative reference information. 
The prior in equation \eqref{eq:sb2} has mean and variance
\begin{align}
    \mathbb{E}[\omega_{jk}] = \frac{\alpha_{jk}}{q_{jk} - 1}\kappa_{jk}, \quad q_{jk}>1, \quad  \text{Var}[\omega_{jk}] = \frac{\alpha_{jk}(\alpha_{jk}+ q_{jk}-1)} {(q_{jk} - 1)^2 (q_{jk} - 2)}\kappa_{jk}^2, \quad{q_{jk}>2}.
\end{align}
If a prior graph is available, say, we have reference edge weights $\omega_{jk}^{(R)}$ from external data, we can center our $\mathrm{beta}2$ prior at those values by setting $q_{jk} > 1$ and  $\kappa_{jk} = \frac{q_{jk}-1}{\alpha_{jk}}\omega_{jk}^{(R)}$ so that $\mathbb{E}[\omega_{jk}] = \omega_{jk}^{(R)}$. The parameter $\alpha_{jk}$ controls behavior near zero: $\alpha_{jk}\le 1$, favors  many small weights, 
increasing $\alpha_{jk}$ 
concentrates mass more tightly around $\omega_{jk}^{(R)}$. $q_{jk}$ controls tail heaviness: smaller $q_{jk}$ yields heavier tails and greater robustness to large edges.
In particular, for a finite variance and tighter concentration one can consider $q_{jk} > 2$ and larger $\alpha_{jk}$. Compared with Gamma or log–normal priors, the $\mathrm{SBeta2}$ prior separates the location (via the scale $\kappa_{jk}$) from the tail index (via $q_{jk}$), enabling centering at $\omega_{jk}^{(R)}$ while independently controlling tolerance for large deviations. Learning $\omega_{jk}$ under the $\mathrm{SBeta2}$ hierarchy, rather than fixing it, affords (i) outcome-specific adaptation, allowing edge weights to adjust to the data, and (ii) uncertainty quantification, since posterior credible intervals for each $\omega_{jk}$ distinguish data-supported connections from those driven primarily by the prior.

We place standard priors on the remaining parameters: 
$\zeta^2 \sim \text{IG}(\frac{1}{2}, \frac{1}{\nu})$, $\nu \sim \text{IG}(\frac{1}{2}, \frac{1}{\zeta_0 ^{2}})$ \citep{makalic2015simple}. 
We specify \(\zeta_0\) according to \cite{piironen2017sparsity}, under the prior assumption that approximately 10\% of the features are expected to have nonzero effects. 
$\tau^2 \sim \text{IG}(a_\tau, b_\tau)$ with $a_\tau > 2$ and 
$b_\tau = 1$, yielding a right-skewed, heavy-tailed 
distribution while ensuring finite variance. 
$\sigma^2 \sim \text{IG}(a_\sigma, b_\sigma)$ with 
$a_\sigma > 2$ to ensure finite variance, and 
{\color{blue}$b_\sigma = (1-R^2)\,\mathrm{Var}(\boldsymbol{y})$}, where $R^2 \in (0,1)$ 
represents the coefficient of determination. 
The parameter $R^2$ should reflect prior expectations about the 
model fit; when such information is not available, a neutral 
choice is $R^2=0.5$.

\subsection{Posterior inference}
We conduct posterior inference with a customized Gibbs sampler that sequentially updates each parameter from its full conditional distribution. For conditionals that lack a closed form, we develop updates based on elliptical slice sampling (ESS), an adaptive approach that works well for multivariate Gaussian priors, enabling sampling from the intractable posterior \citep{murray2010elliptical}. In this section, we provide an overview of the computational challenges addressed by our Gibbs sampling algorithm. Complete derivations of the Gibbs updates for all the parameters can be found in Section 3 of the Supplementary Material.

\paragraph{Posterior conditional of \(\bm{\psi}\)}

The log posterior conditional distribution of \(\bm{\psi}\), given the prior and likelihood, is:
\[
\log p(\bm{\psi} \mid \cdot) \propto 
-\frac{1}{2\sigma^2\zeta^2} \sum_{j=1}^p e^{-2\psi_j} \left[ (\mathbf{U}\bm{\beta})_j \right]^2 
-\frac{1}{2} \sum_{j < k} \frac{1}{\tau^2}\,\omega_{jk}\, (\psi_j - \psi_k)^2 
-\sum_{j=1}^p \psi_j
\]

Because the posterior is intractable, we update \(\boldsymbol{\psi}\) using a hybrid elliptical slice sampling (ESS) scheme. ESS generates proposals along ellipses determined by the Gaussian prior covariance and adaptively selects the step size by slice sampling, thereby respecting posterior geometry without tuning parameters \citep{murray2010elliptical}. To improve mixing when the likelihood induces sharp local curvature, we periodically replace the ESS move, after every \(K\) iterations, with coordinate-wise one-dimensional slice updates for each component of \(\boldsymbol{\psi}\) \citep{neal2003slice}. This hybrid ESS + slice sampler combines global moves with local moves, yielding robust exploration of the full conditional posterior of \(\boldsymbol{\psi}\). A technical description is provided in Algorithm S1 of Supplementary Section 4.

\paragraph{Posterior conditional of $\omega_{jk}$}We place the scaled beta2 prior
$\omega_{jk} \sim \mathrm{SBeta2}(\alpha_{jk}, q_{jk}, \kappa_{jk})$, which admits a
gamma-mixture representation  through an auxiliary variable
$\rho_{jk}$ \citep{perez2017scaled}. Combining this representation with the likelihood contribution of
$\omega_{jk}$ yields closed-form full conditionals for both $\omega_{jk}$ and
$\rho_{jk}$.
Therefore, the full conditional distribution of \(\omega_{jk}\) and  \(\rho_{jk}\) is
\begin{align*}
\omega_{jk}\mid \rho_{jk},\psi_j,\psi_k,{\color{blue}\tau^2}
\sim
\text{Gamma}\left(
\alpha_{jk},
\frac{\rho_{jk}}{\kappa_{jk}}
+
\frac{(\psi_j-\psi_k)^2}{2\tau^2}
\right).\\
p(\rho_{jk}\mid \omega_{jk})
\propto
\rho_{jk}^{\alpha_{jk}+q_{jk}-1}
\exp\left\{
-\left(1+\frac{\omega_{jk}}{\kappa_{jk}}\right)\rho_{jk}
\right\}
\end{align*}
Introducing $\rho_{jk}$ thus avoids direct sampling from the
marginal scaled beta2 posterior. We collect the edge weights into a symmetric
matrix $\mathbf{W}$ with $\omega_{jk} = \omega_{kj} \geq 0$ and $\omega_{jj} = 0$, and
resolve the scale non-identifiability between $\mathbf{W}$ and $\tau^2$ by
row-normalizing $\mathbf{W}$.
\subsection{Feature selection and  evaluation of predictive performance}
\label{ssec:predaccuracy}
To identify a subset of informative predictors, given posterior samples $\{\psi_j^{(s)},\,\zeta^{2\,(s)}\}_{s=1}^{S}$, for each draw $s$, we compute shrinkage weights
$
\kappa_j^{(s)} \;=\; \frac{1}{\,1 \;+\; n\exp(2\psi_j^{(s)}) \,\zeta^{2\,(s)}}\,,
$
which lies in the interval $(0,1)$.  
We summarize the posterior mean shrinkage weight,
$
\widehat{\kappa}_j \;=\; \frac{1}{S}\sum_{s=1}^{S} \kappa_j^{(s)} \, 
$ and 
declare feature $j$ to be active/selected whenever 
$
(1-\widehat{\kappa}_j) \;\geq\; 0.5
$ \citep{carvalho2010horseshoe}. 
Predictive performance is assessed using the prediction error (RMSE) between 
observed outcomes $\boldsymbol{y}$ and their fitted values
$\hat{\boldsymbol{y}} = \mathbf{X}\hat{\boldsymbol{\beta}}$, and the estimation error
$\lVert \hat{\boldsymbol{\beta}} - \boldsymbol{\beta}_{\mathrm{true}} \rVert_2$. Variable-selection accuracy is aummarized using Matthews correlation coefficient (MCC),
which gives a balanced score from the true/false positive and negative counts and is
robust to signal imbalance \citep{chicco2020advantages}.


\subsection{Graph recovery}
\label{ssec: GraphRecovery}
For network recovery, we summarize the posterior estimates of the edge weights \(\{\omega_{jk}\}_{j,k=1}^p\) using a row-stochastic weight matrix \(\mathbf{W} \in \mathbb{R}^{p\times p}\), defined entry-wise as
\(\mathbf{W}_{jk} = \frac{\omega_{jk}}{\sum_{\ell \neq j}\omega_{j\ell}}\,\mathbf{1}\{j \neq k\}.
\)
By construction, each row of \(\mathbf{W}\) sums to unity.
Consequently, \(\mathbf{W}_{jk}\) can be interpreted as the proportion of node \(j\)’s total outgoing connection strength allocated to node \(k\).  
In order to convert the continuous posterior weights  $\{\omega_{jk}\}$ into a sparse, interpretable graph, we first apply a magnitude floor of $\varepsilon = 10^{-3}$, thereby discarding any link that carries less than 0.1\,\% of a node’s total outgoing influence in the densest possible setting ($\bar w = 1/(p-1)$).  For each unordered pair $(j,k)$ we compute the posterior inclusion probability
\(
\mathrm{PIP}_{jk} \;=\; \frac{1}{S}\sum_{s=1}^S \mathbf{1}\bigl(|\omega_{jk}^{(s)}| > \varepsilon\bigr)\,,
\)
where $S$ is the number of post–burn-in MCMC draws.  We construct the median‐probability estimated graph by retaining every undirected edge $(j,k)$ which  satisfies
$
\mathrm{PIP}_{jk} \;\ge\; 0.5
$.
%
Recovery is evaluated against the true block structure with three complementary
criteria. We report PIP-Block and PIP-Cross, the average PIP within and across true coefficient blocks, and the MCC of the
median-probability graph. Treating $\{\mathrm{PIP}_{jk}\}$ as continuous scores
against the binary labels $l_{jk} = \mathbf{1}\{j,k \text{ in the same true block}\}$,
we also compute the AUROC and AUPR using the \texttt{PRROC} package. Finally, we apply
the Louvain algorithm to the thresholded true and estimated graphs and compare the
resulting community assignments by the adjusted Rand index (ARI).

\section{Simulation Study}
 \subsection{Simulation setup}
\label{sec:sim-setup}

We design a simulation study to evaluate recovery of the true similarity graph, identification of true covariate effects, and estimation accuracy for the coefficients of the compositional predictors. The data-generating mechanism proceeds in five steps. Here, we provide a concise summary; full details, including parameter values, the coefficient-smoothing prior, and the graph-construction formulas, are provided in Supplementary Section 6.

We define two latent signal
clusters, $C_1 = \{18,\dots,23\}$ and $C_2 = \{40,\dots,45\}$, and construct a true
similarity graph $\mathbf{W}$ whose entries decay slowly within clusters and rapidly
in the background. This yields strong within-cluster connectivity and weak
between-cluster similarity, so that $\mathbf{W}$ encodes two well-separated functional
modules. Signal magnitudes are smoothed over $\mathbf{W}$ through a
graph-Laplacian Gaussian prior, so that the
entries of the coefficient vector $\boldsymbol{\beta}$ vary smoothly across connected
features. Cluster $C_1$ receives positive and $C_2$ negative effects, introducing
contrast between the two modules; the active coefficients are centered to sum to zero,
and all remaining $\beta_j = 0$.

To assess sensitivity to realistic graph
misspecification, we generate three perturbed versions $\widetilde{\mathbf{W}}$ of the
true graph $\mathbf{W}$, each used as the prior similarity matrix supplied to the
graph-aware methods. The \emph{high-overlap} graph is a conservative within-band
subgraph of $\mathbf{W}$ with no false positives that recovers
roughly half of the true edges. The \emph{low-overlap} graph retains about $20\%$ of
the true edges and adds substantial false-positive edges. The \emph{hub-structure}
graph concentrates the perturbation around a subset of high-degree nodes, partially
preserving the backbone while adding and dropping edges near hubs. Construction
details and a quantitative characterization of each regime are given in Supplementary
Section~6.

Next, we draw a latent Gaussian design
from a logistic-normal model with AR(1) baseline
with an additional covariance boost of on every pair
connected in $\widetilde{\mathbf{W}}$.The boost ensures a realistic and moderate increase in edge correlations. The latent
draws are exponentiated, normalized to the simplex, and log-transformed to give the
compositional predictor matrix $\mathbf{X} \in \mathbb{R}^{n\times p}$, and the outcome
is $y_i = \boldsymbol{x}_i^\top \boldsymbol{\beta} + \varepsilon_i$ with
$\varepsilon_i \sim N(0,\sigma_y^2)$, and $\sigma_y^2 = 1.6$, where
$\boldsymbol{x}_i$ denotes the $i$-th row of $\mathbf{X}$, yielding a moderate
signal-to-noise ratio.

To mimic real microbiome data, counts are
generated from the latent compositions using negative-binomial library sizes and a
Dirichlet-multinomial sampling model, with a detection limit that introduces sampling
zeros. Varying the count parameters yields zero-inflation rates ranging from $10\%$ to
$40\%$ (zi10, zi20, zi30, zi40) while holding the latent compositions and covariance
structure fixed. For downstream analysis we add the sample-specific pseudo-count of
\citep{shi2022high} to each feature and apply a CLR transformation, producing a noisy
predictor matrix $\mathbf{X}$, that we use to assess robustness to count-level noise.

\vspace{-0.4cm}\subsection{Benchmarking methods}
\label{ssec:Benchmarking Methods}
\noindent We benchmark the performance of our method against several existing approaches. The first is 
 \textbf{generalized CLR lasso}, which applies generalized lasso regression \citep{ali2019generalized} to centered log-ratio (CLR) transformed compositional covariates. To incorporate graph information, we construct a penalty matrix 
from the supplied graph \(\widetilde{\mathbf{W}}\). Specifically, we define
\(\mathcal{E}
=
\{(j,k): 1 \leq j < k \leq p,\; \widetilde{\mathbf{W}}_{jk} > 0\}
=
\{(j_\ell,k_\ell)\}_{\ell=1}^m\).
The penalty matrix \(\mathbf{D}\) is then given by
\(\mathbf{D}_{\ell j}
=
\sqrt{\widetilde{\mathbf{W}}_{j_\ell k_\ell}}\,
\mathbf{1}\{j=j_\ell\}
-
\sqrt{\widetilde{\mathbf{W}}_{j_\ell k_\ell}}\,
\mathbf{1}\{j=k_\ell\}\),
so that
\(\|\mathbf{D}\boldsymbol{\beta}\|_1
=
\sum_{\ell=1}^m
\sqrt{\widetilde{\mathbf{W}}_{j_\ell k_\ell}}\,
|\beta_{j_\ell}-\beta_{k_\ell}|\). We also compared against \textbf{CODA lasso} \citep{lin2014variable}, \textbf{BAZE} \citep{zhang2021bayesian}, and \textbf{BCGLM} \citep{zhang2024bayesian}, as described in Section~\ref{sec:Introduction}.
We used default hyperparameter settings for the Bayesian models and supplied the misspecified similarity matrix \(\widetilde{\mathbf{W}}\) as prior graph information for BAZE, BCGLM, and GRACE.

For data sets that mimic zero inflation, we consider $n = 400$ and $p=300$ with rates of zero inflation ranging from 10-40\%. Relative abundance data without sampling noise are analyzed for $p=100, 300$ and $n = 400$. For each \((n, p)\) configuration, we randomly split the data into equal-sized training and test sets, and report the prediction accuracy and graph recovery metrics described in Sections \ref{ssec:predaccuracy} and \ref{ssec: GraphRecovery} averaged over 30 independent replicates.

\section{Results}
We benchmark the prediction and network recovery performance of GRACE against the alternative methods across three graph structures: high, low, and hub structure. For purposes of brevity, we only include results for datasets that mimic zero-inflation in the main manuscript as it more strongly resembles real microbiome data. The analysis for the relative abundance data, without zero inflation, is presented in Supplementary Section \textcolor{blue}{8.} 


\subsection{Prediction evaluation}
\begin{figure}
    \centering
    \includegraphics[width=\linewidth]{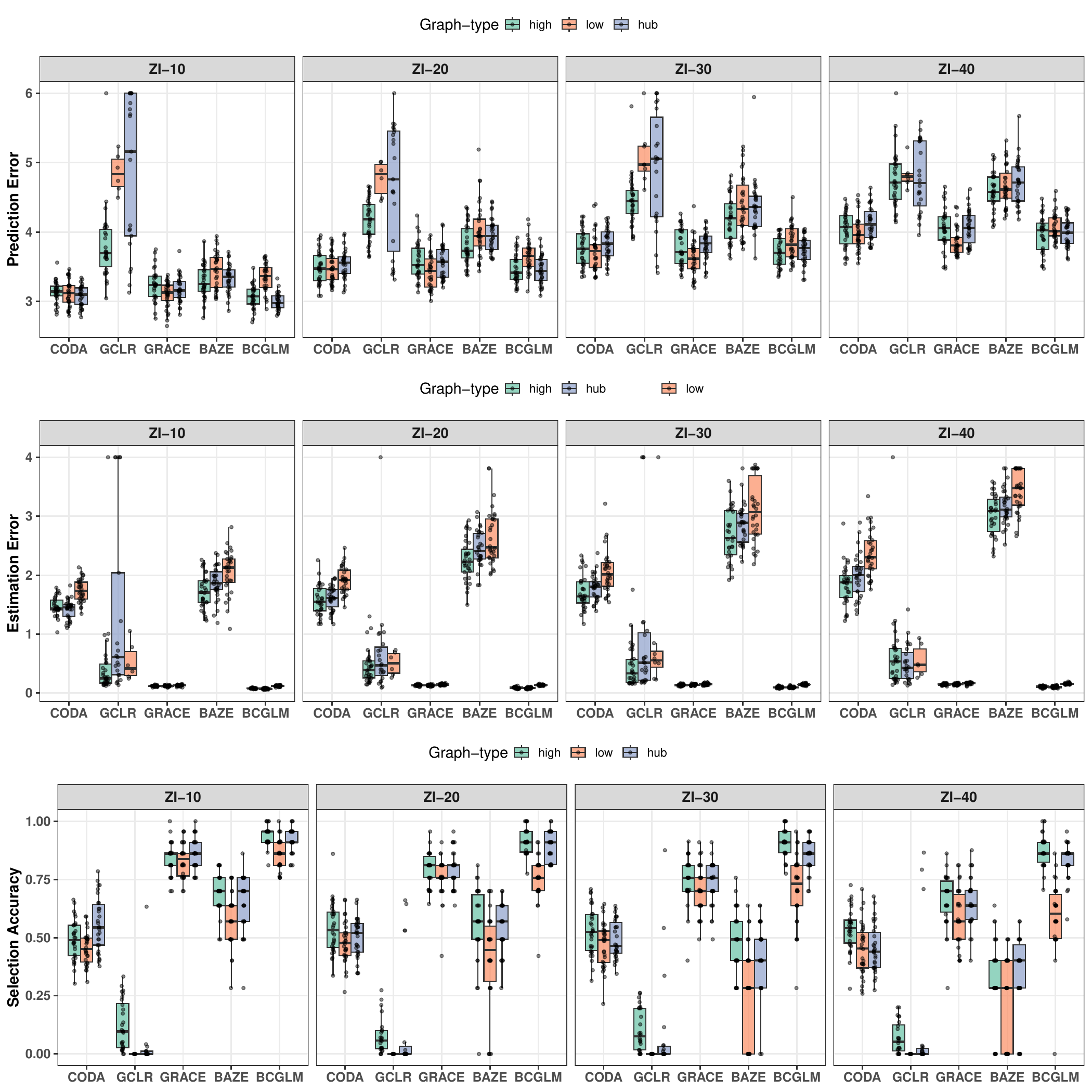}
    \caption{Performance comparison of GRACE and the benchmarking methods  across varying levels of zero inflation (ZI = 10–40) and graph types (high, low, hub structure). Boxplots show the distribution of prediction error, estimation error, and selection accuracy over 30 replicates. GRACE demonstrates stable selection error and competitive prediction error across settings, while also achieving superior variable selection accuracy, particularly when zero inflation is moderate.}
    \label{fig:ZI.metrics_byScenario}
\end{figure}

Figure \ref{fig:ZI.metrics_byScenario} illustrates the  performance of GRACE along with the benchmarking methods in terms of prediction error, coefficient estimation error, and MCC of feature selection for simulated datasets with rates of zero inflation ranging from 10-40\%. BCGLM and GRACE are the best performing approaches in all metrics, with BAZE having a better MCC than the other two frequentist approaches. There is an expected decrease in prediction accuracy as zero inflation of the covariates increases.  Despite this global trend, GRACE exhibits tight interquartile ranges and stable medians, indicating robustness to excess zeros. The results for the metrics vary with the structure of the underlying graph, with the high and hub structure settings demonstrating consistently better performance for GRACE across all metrics. 
When the true signal graph $\mathbf{W}$ aligns with the design graph $\widetilde{\mathbf{W}}$ (high and \textcolor{blue}{hub structure}, Supplementary Section 6, Table S3, S4 and Figure S11), the signal to noise ratio improves, yielding better prediction  metrics, especially for graph-aware methods like GRACE. Under low overlap, misaligned correlations  affect the method performance. The impact of graph alignment is further amplified due to zero inflation in the observed data.

\subsection{Graph recovery}
\begin{figure}
    \centering
    \includegraphics[width=\linewidth]{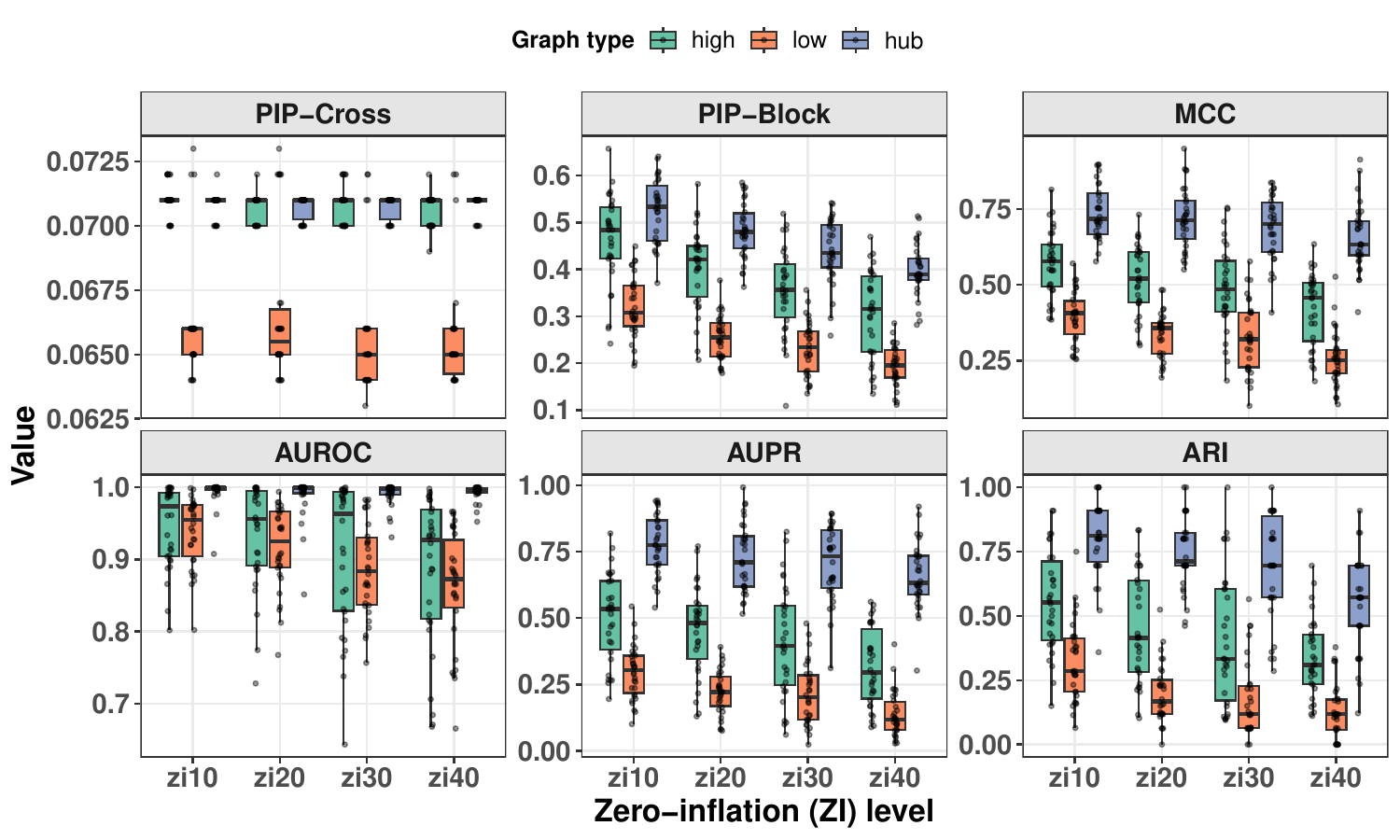}
    \caption{Effect of zero inflation on edge recovery. Metrics are shown versus the zero-inflation level (zi10–zi40) for the same three graph designs. AUPR and MCC decline with increasing zeros, while AUROC is more stable; PIP patterns reflect diminishing within-block evidence and occasional cross-block inflation under severe ZI. hub structure is most sensitive; high-overlap remains comparatively robust.}
    \label{fig:pA_with_hub}
\end{figure}

Next, we highlight a key novel feature of GRACE: its ability to estimate a graph adapted to the signals in the data.
To highlight this inferential result of our proposed model, we now evaluate the quality of the graphs estimated by GRACE. Since none of the other benchmarking methods produce graph outputs, we report the edge recovery results of GRACE across three graph designs, for the simulation datasets with increasing levels of zero inflation. As shown in Figure \ref{fig:pA_with_hub},  recovery performance generally declines with increasing zero inflation. PIP-Block, AUPR, and MCC are the most sensitive, deteriorating
sharply in the low-overlap setting, whereas AUROC remains comparatively robust
and PIP-Cross is largely stable, indicating persistent cross-block inflation
under low overlap. The hub-structure design is the most resilient under severe
zero inflation, high-overlap offers moderate robustness, and low-overlap degrades
substantially. These differences trace to how each perturbation propagates from the observed
predictor graph to the recovered community structure, examined in detail in
Supplementary Section~6 (Figure~S11; Tables~S3--S4). In brief, the hub-structure design preserves true edges around high-degree vertices
with balanced short- and long-range recall and the strongest community structure,
making graph recovery most resilient as zero inflation increases; the high-overlap
design captures only local edges (high precision, no long-range recovery); and the
low-overlap design adds cross-block links that degrade both precision and recall
(Supplementary Section~6, Figure~S11, Tables~S3--S4) The relative-abundance
results without sampling noise follow the same pattern (Supplementary
Section~8, Figure~S10).

In summary, our results indicate that out-of-sample prediction remains
robust even when the predictor graph is misspecified, whereas graph
recovery degrades substantially when encoded feature relationships
diverge from those driving the outcome. 
\textcolor{blue}{Our prior
sensitivity analysis (Supplementary Section~7) offers practical guidance
in this setting: when the supplied graph is well aligned with the
outcome-relevant structure, a strongly informative graph prior improves
edge recovery, whereas when the external graph is uncertain or likely
misspecified, a weakly informative prior allows the posterior graph to
be learned primarily from the data and yields better recovery of the
underlying outcome-driven modules. We therefore recommend matching the
strength of the graph prior to one's confidence in the external graph,
and defaulting to a weak prior when substantial divergence is
suspected.} Overall, our results highlight the potential of our new
approach to characterizing relations among the predictors in
compositional regression.



\section{Application to oral microbiome data}
\label{sec:case_study}

In this section, we analyze the ORIGINS data decribed in Section \ref{sec:Introduction} using GRACE as well as the competing models described in Section \ref{ssec:Benchmarking Methods}. The dataset was  preprocessed by removing repeated measurements and samples with missing insulin and glucose values. Rare taxa were then excluded by filtering out features present in fewer than 30\% of samples. 
\textcolor{blue}{This threshold was chosen to remove highly sparse taxa while retaining a representative feature set: at the 30\% threshold, the median retained taxon was present in a majority of samples and the retained taxa captured 97.7\% of the total observed abundance.}
We removed low-coverage samples by calculating sequencing depth per sample and excluding those below the 5th percentile of the depth distribution. After preprocessing, the final dataset comprised 62 taxa across 114 samples. 
The prior graph was derived from the available phylogenetic tree by connecting each taxon to its 
five nearest neighbors based on phylogenetic covariance, and then symmetrized to form an undirected adjacency matrix (Figure S8). To handle zeros in the abundance matrix \(\mathbf{M}\), we apply the variable-correction transform \citep{shi2022high}, as described in detail in Supplementary Section 9.1.


\subsection{Prediction and variable selection} \label{subsec::Prediction and selection results}
\begin{figure}[htbp]
    \centering
    \begin{subfigure}[t]{\linewidth}
        \centering
        \includegraphics[width=\linewidth]{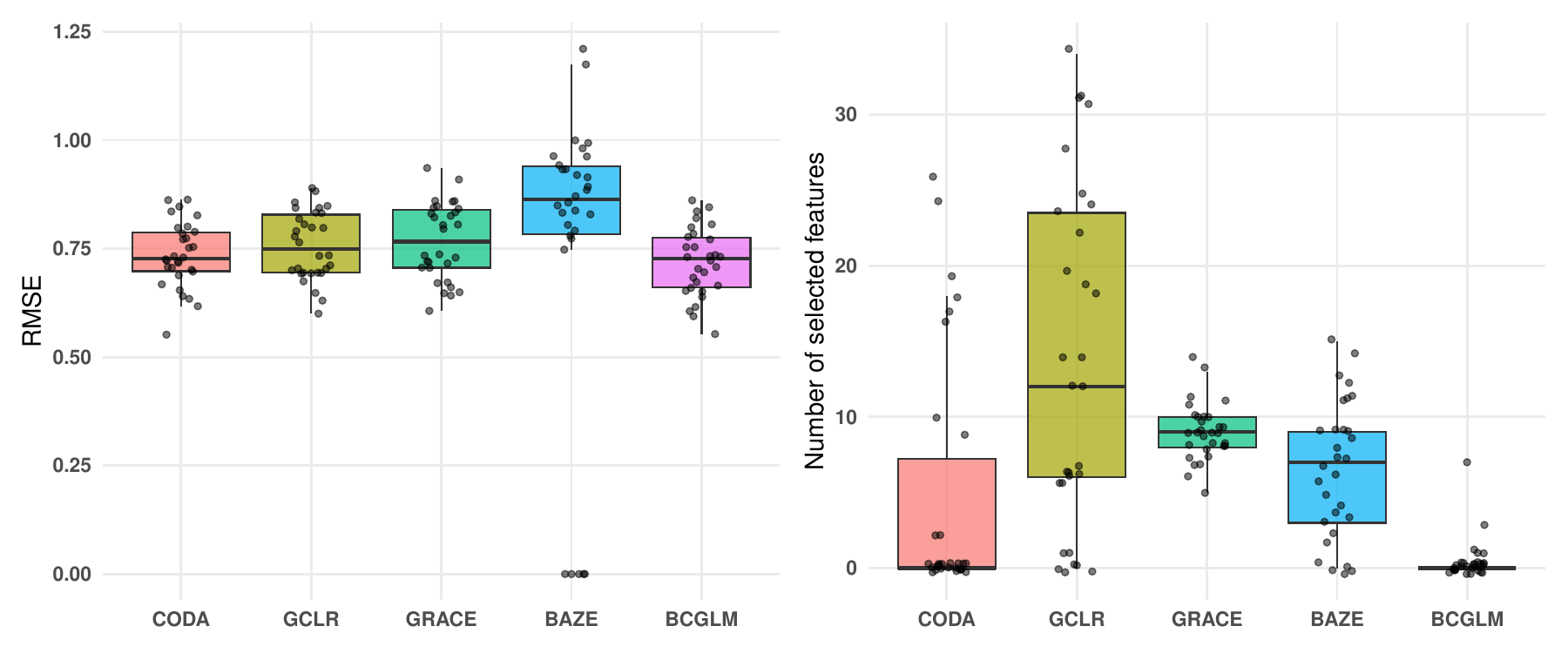}
        \caption{Comparison of model performance and feature selection across methods for the real data. 
        (Left) Distribution of root mean squared error (RMSE) values over 30 data splits for each method. Lower RMSE indicates better predictive accuracy. 
        (Right) Distribution of the number of selected features for each method over the same splits. Higher counts indicate less sparsity in feature selection.}
        \label{fig:Real_Data_Splits}
    \end{subfigure}
    \vspace{1em} 
    \begin{subfigure}[t]{\linewidth}
        \centering
        \includegraphics[width=\linewidth]{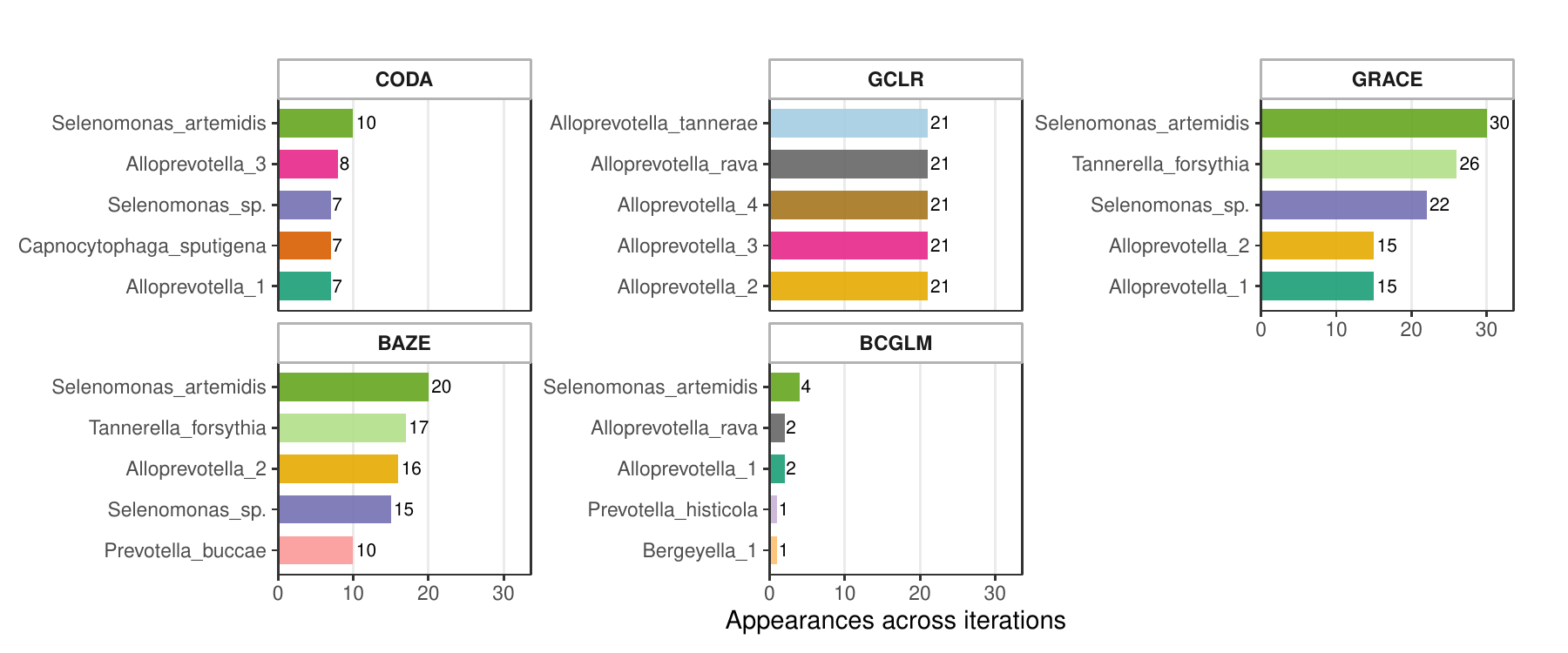}
        \caption{Top five taxa most frequently selected across iterations for each method. Bars represent the count of iterations a taxon was selected.}
        \label{fig:Top_Taxa}
    \end{subfigure}
    \caption{Summary of prediction performance, variable selection, and most frequently selected taxa across methods for the real data analysis.}
    \label{fig:RealDataSummary}
\end{figure}

We randomly divided the 114 samples into a training set of 79 samples and a test set of 35 samples, and fit the  models on the training data. The analysis was repeated over 30 independent replicates, with fitted models used to evaluate predictive performance on the corresponding test sets. The resulting distributions of prediction errors across all methods are presented in Figure~\ref{fig:RealDataSummary}. Among the competing approaches, GRACE achieved competitive predictive accuracy while consistently selecting similar number of features across replicates. Notably, \textsl{Selenomonas artemidis} was identified by all methods except GCLR. Furthermore, BAZE and GRACE exhibited closely aligned feature selection patterns, sharing four of the five most frequently selected features. 
This sample-splitting analysis demonstrates the stability of the selected associations across repeated splits while maintaining competitive predictive accuracy, consistent with the simulation results.

To further investigate the influence of oral microbiome composition on insulin regulation, we applied the proposed model to the complete dataset. Table S6 lists the six taxa with median regression coefficients $\lvert\beta_j\rvert$ exceeding 0.1. Among these, \textsl{Tannerella forsythia} and \textsl{Selenomonas artemidis} were selected based on their posterior shrinkage weight quantiles. For both species, the 95\% credible intervals for $\beta_j$ excluded zero. Among these species, \textsl{Tannerella forsythia} 
has long been recognized as one of the bacteria that contribute to periodontitis \citep{Socr1998}.  More recently, increased abundance of \textsl{T.\ forsythia} in the oral microbiome has been associated with higher fasting blood glucose levels \citep{Chang2023}. Previous studies have reported a relatively higher abundance of \textsl{S. artemidis} in the salivary microbiome of untreated diabetic patients compared to both healthy individuals and diabetic patients receiving treatment \citep{yang2020changes}. 
\textcolor{blue}{To assess robustness to potential confounding, we repeated the analysis with adjustment for age, sex, race, and BMI. The results were consistent with the primary analysis: BMI was the only covariate with evidence of a nonzero association, and the leading positive-effect taxa \textsl{T. forsythia} and \textsl{S. artemedis} remained selected (Supplementary Section~10, Tables S1, S2, Figures S6, S7).}



\subsection{Similarity graph recovery} \noindent


To summarize posterior edge evidence, we computed PIPs for each taxon--taxon pair. 
At each MCMC iteration $s$, we formed the row-normalized off-diagonal precision matrix $\widehat{\mathbf{W}}^{(s)}$, vectorized its upper triangle, and recorded indicators $\mathbf{1}(|w_\ell|>\varepsilon)$ with $\varepsilon=10^{-3}$. 
The PIP for edge $\ell$ was estimated as the post--burn-in proportion of samples with $|w_\ell|>\varepsilon$ and averaged across chains. 
A weighted taxon network was then constructed using these PIPs, retaining only high-confidence edges above the 90th percentile.
\textcolor{blue}{This percentile-based rule is scale-adaptive and controls the density of the posterior graph summary. For \(p=62\) taxa, retaining the top \(10\%\) undirected edges corresponds to an average degree of about 6, comparable to the five-nearest-neighbor phylogenetic reference graph used in the real-data prior.} Next, we partitioned graph nodes by the sign of their posterior median regression coefficients \(\hat{\beta}_j\), forming positive and negative subgraphs. Louvain community detection was then applied separately within each subgraph to identify taxa that are both strongly connected through edges with high posterior support and associated with the outcome in the same direction.



\begin{figure}
    \centering
    \includegraphics[width =\linewidth]{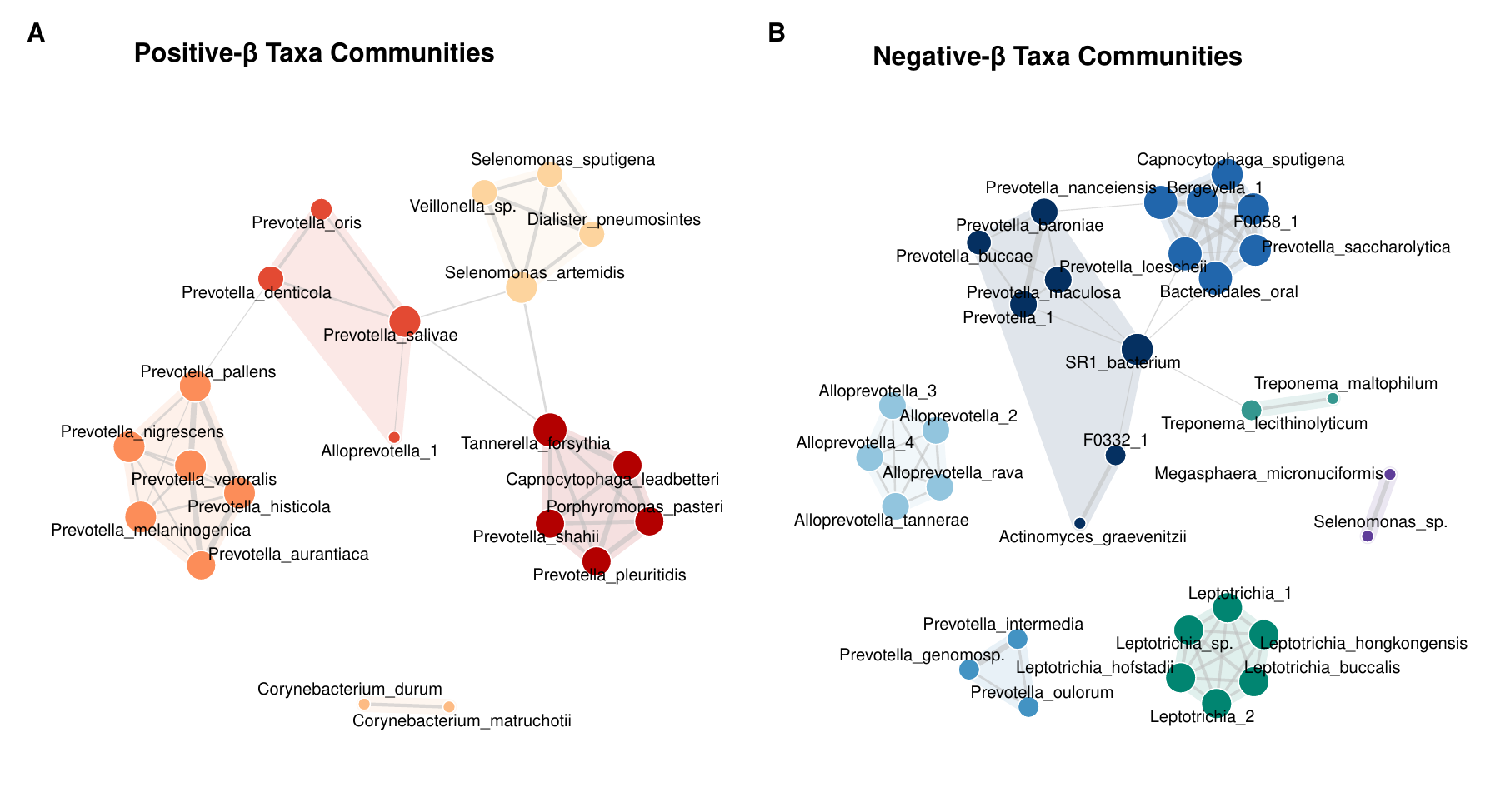}
    \caption{Two‐panel network visualization of microbial taxa with positive (left) and negative (right) associations. Each node represents a taxon whose  regression coefficient is positive or negative, respectively; node color indicates its community membership and node size scales with its degree. Edges show the top 10 \% of posterior inclusion probabilities (PIP) for taxa–taxa associations estimated by the model, with line width proportional to PIP; taxa without any such high-confidence edges were omitted.}
    \label{fig:real_network_moderate}
\end{figure}

\textcolor{blue}{Figure~\ref{fig:real_network_moderate} shows the posterior taxa subgraphs under the moderate prior. The positive-effect network includes a Selenomonas-rich module containing the selected taxon \textit{Selenomonas artemidis}, and a second module involving \textit{Tannerella forsythia}, which was selected by the prediction model, along with \textit{Porphyromonas pasteri} and \textit{Capnocytophaga leadbetteri}, which have previously been associated with periodontitis \citep{relvas2021relationship,lenartova2021oral}. The negative-effect network includes Prevotella/Bacteroidales and Alloprevotella modules. Thus, the learned graph organizes taxa into outcome-relevant effect-sharing modules rather than only producing a list of selected taxa.}



\textcolor{blue}{Next, we assessed the robustness and interpretation of these
graph summaries. To evaluate sensitivity
to the strength of the edge-weight prior, we considered strong,
moderate, and weak settings (Supplementary Section~9.2); the leading
outcome-associated taxa remained graph-connected across all prior
specifications, and their posterior coefficient magnitudes varied by
less than $10\%$ (Figure~S2), indicating that the main findings are not
artifacts of a particular prior choice. The recovered communities were
not redundant with a clustering of the signed regression coefficients
(Supplementary Section~9.4, Figure~S5), and a literature-based
enrichment analysis indicated that the strongest positive-effect
community was enriched for taxa with prior periodontal and metabolic
relevance (Supplementary Section~9.3, Figures~S3--S4). Overall, GRACE identifies outcome-predictive features with independent biological support and provides an outcome-driven graph summarizing their relationships. This graph should be interpreted as a hypothesis-generating summary of outcome-driven relationships rather than a taxa–taxa interaction network.
}

\section{Discussion}
\label{sec:discuss}

\textcolor{blue}{GRACE provides a unified framework for variable selection and outcome-informed graph learning in microbiome compositional regression. By using prior network information as a guide while allowing edge weights to adapt to the data, GRACE estimates a shrinkage graph linking taxon regression effects, rather than assuming that a fixed phylogenetic or taxonomic graph captures the relevant effect-sharing structure.
}

\textcolor{blue}{Our approach connects to the broader literature on structured estimation. Network-constrained regularization \citep{li2008network}  impose Laplacian penalties directly on the regression coefficients, encouraging graph-connected features to have similar signed effects. However, these methods are not designed for compositional predictors and typically rely on a fixed graph. In contrast, GRACE enforces the compositionality constraint while learning outcome-informed edge weights that adapt the shrinkage structure to the data. Fixed-graph coefficient-smoothing approaches are represented in our benchmarks by GCLR, BAZE and BCGLM. Other related methods include adaptive gPCA \citep{fukuyama2019adaptive}, which also learns the degree of reliance on phylogenetic structure, but in an unsupervised dimension-reduction setting, and 
generalized matrix decomposition regression \citep{wang2023generalized}, which exploits two-way structure through low-rank decomposition, but does not impose compositional constraints or produce an estimated outcome-driven feature graph.}

Across simulations and the real data application, GRACE achieved competitive prediction and feature-selection performance while showing that outcome-relevant predictor relationships can differ substantially from evolutionary or taxa-distance-based graphs. \textcolor{blue}{Because the learned graph is defined through the shrinkage structure rather than through the regression coefficients themselves, it should not be interpreted as a microbial interaction network. Instead, it provides a hypothesis-generating summary of outcome-driven similarity among taxa. In ORIGINS, the recovered communities were not simply clusters of signed posterior coefficients, and literature-based enrichment suggested that the strongest positive-effect community contained taxa with prior periodontal and metabolic support.}

To improve computational scalability, we developed a hybrid Gibbs sampler with elliptical slice sampling. In practice, preprocessing to remove highly sparse features remains important to reduce instability from pseudocount adjustments. Future work could extend GRACE by directly modeling zero-inflated counts and by incorporating multiple noisy network sources to improve robustness beyond the single-source prior setting considered here.

\backmatter
%
%
\section*{Acknowledgements}
This research was partially supported by NIH R01 HL158796.



\section*{Supplementary Materials}

Supplementary material, including figures and tables and code for implementing the model, is available with this paper at the Biometrics website on Oxford Academic.

\section*{Data availability}
The fastQ files for our case study can be accessed from  the European Nucleotide Archive using project ID PRJEB50306. The metadata and processed sequences can be downloaded from Qiita using study ID 11808. 


\vspace*{-8pt}



\bibliography{biomsample.bib}
\bibliographystyle{plainnat}

\appendix




\label{lastpage}

\end{document}